\documentclass[twocolumn]{revtex4-1}
\usepackage{amssymb}
\usepackage{amsfonts}
\usepackage{amsmath}%
\usepackage{graphicx}
\usepackage{hyperref}
\usepackage{color}

\begin{document}

\title{Angle dependent conductance in graphene}
\author{C. H. Fuentevilla}
\author{J.D. Lejarreta}
\affiliation{Departamento de F\'{\i}sica Aplicada, Universidad de Salamanca, E-37008 Salamanca, Spain}
\author{C. Cobaleda}
\author{E. Diez}
\affiliation{Laboratorio de Bajas Temperaturas, Universidad de Salamanca, E-37008 Salamanca, Spain}

\begin{abstract}

In this paper, we study a theoretical method to calculate the conductance across a square barrier potential in monolayer graphene. 
We have obtained an analytical expression for the transmission coefficient across a potential barrier for monolayer graphene. 
Using the transmission coefficient obtained we have an analytical expression for the conductance. This expression will be used to calculate the conductance in the case in which there is a potential barrier, which in our case will modelise the behaviour of a top gate voltage of a field effect transistor. Once this analysis has been performed we study the scenario in which carriers scatter with the potential barrier with different incidence angles and we have found that for any incident angle an effective gap is induced.

\end{abstract}
\maketitle

\section{Introduction}

Monolayer graphene is a one atom thick carbon layer in a hexagonal honeycomb lattice. From the lattice properties one can deduce the band structure and, hence, the energy spectrum.

In graphene, charge carriers are described by the  Dirac equation instead of Schr\"odinger's as it is the case of traditional semiconductors. This is a direct consequence from the fact that graphene has two equivalent triangular sublattices A and B\cite{Castro_Neto}. 
Therefore, the dispersion relationship is linear and is expressed by $E=\pm \hbar v_F k$ \cite{Katsnelson1} where the positive (negative) sign describes electrons (holes) as carriers. Thus, it is deduced that carriers in graphene have the same dispersion relationship than massless particles with velocity $v_F\sim c/300$. 

According to previous studies, graphene might be the ideal material for electronic devices\cite{Lemme,Lin,Xia} due to several properties which are present in graphene such as its high mobility (up to 15000$cm^2/Vs$\cite{Geim1}), the large scattering length\cite{Novoselov} and also because graphene can stand a current density  which is six order of magnitude greater than in copper\cite{Geim}.
The conductance of PN junctions in graphene has been studied previously\cite{Low1} and different configurations of junctions NNN, NPN, PPP and PNP have been also studied\cite{Huard}. Experimental measurements have been carried out in order to study the conducting properties of PNP structures\cite{Gorbachev} obtained by the deposition of a top gate separated from the graphene by an air gap. 
Quantum oscillations of the conductance in graphene have been studied both theoretically\cite{Yampolskii} and experimentally\cite{Young, Begliarbekov}.\\

Because of its electrical properties, graphene is an interesting potential material to develop nanodevices usable in technological applications, such as field effect transistors (FET). 
In this work we have modelised a FET based on graphene and analyzed its electrical transport capabilities based on an exact analytical solution to the Dirac equation. Finally, we study the different behaviour of the conductance shown by a FET considering that the current does not flow perpendicularly to the top gate.

\section{Model}

We start our model considering that the carriers in graphene can be considered as massless ultrarelativistic Dirac particles, since the tight binding hamiltonian for graphene  leads to a linear dispersion relationship. Therefore,  graphene carriers are described by the Dirac eigenvalue equations:
\begin{eqnarray} 
v_{F}{\bf \sigma} \cdot{\bf p}\Psi({\bf r}) = E\Psi({\bf r})
\end{eqnarray}
in where ${\bf \sigma} = (\sigma_{x},\sigma_{y})$ are the Pauli matrices, and $\Psi$ is a two component spinor.
\begin{displaymath}
\sigma_{x}=\left( \begin{array}{clcr}
0&1\\
1&0\end{array} \right),
\sigma_{y}=\left( \begin{array}{clcr}
0&-i\\
i&0\end{array} \right),
\Psi(x,y)=\left( \begin{array}{clcr}
 \varphi_1 (x,y)\\
  \varphi_2 (x,y)\end{array} \right)
\end{displaymath}
Considering an external potencial V(x,y), the hamiltonian is:
\begin{eqnarray}
H=v_{F}( \sigma_x p_x+\sigma_y p_y)+V(x,y)
\end{eqnarray} 
which yields
\begin{eqnarray}
-i \hbar v_F (\frac{\partial \varphi_2}{\partial x}-i\frac{\partial \varphi_2}{\partial y})+V \varphi_1 = E \varphi_1\nonumber\\
-i \hbar v_F (\frac{\partial \varphi_1}{\partial x}+i\frac{\partial \varphi_1}{\partial y})+V \varphi_2 = E \varphi_2
\end{eqnarray} 
In the particular case of V a constant, eventually zero, the solution will be a plane wave spinor in the form of
\begin{eqnarray}\label{wfunction}
\Psi_{\pm}=\frac{1}{\sqrt{2}}e^{i {\bf k}\cdot {\bf r}}\left( \begin{array}{clcr}
1\\
\pm \theta_k\end{array} \right)
\end{eqnarray} 
where 
\begin{eqnarray}
 \theta_k=\arctan\left(\frac{k_y}{k_x}\right)
\end{eqnarray} 
Where the positive (negative) sign will be due to electronlike (holelike) regime.  

\begin{figure}
\centerline{\includegraphics[width = 8.5cm,clip=]{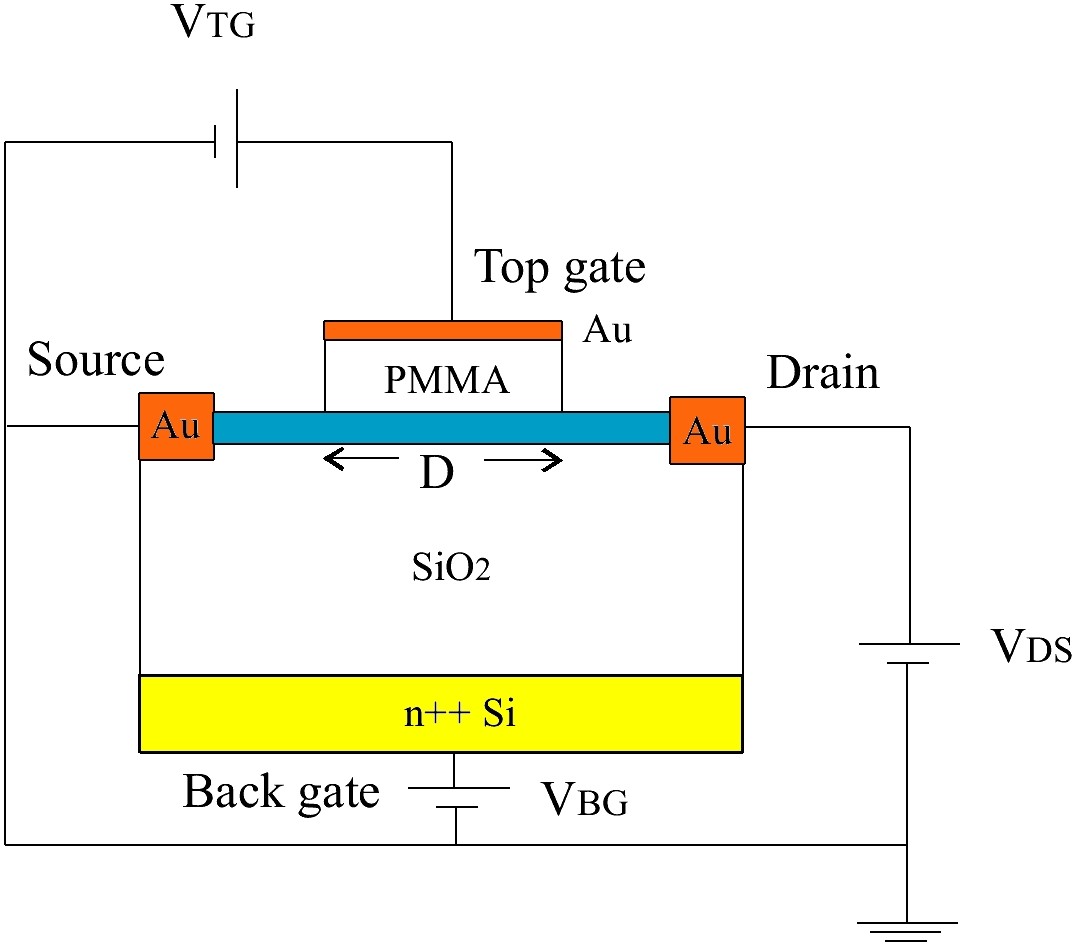}}
\caption{Scheme of the device studied in this paper}
\label{c3}
\end{figure}

Fig.~\ref{c3} shows an scheme of the device under consideration: a rectangular single layer of graphene (blue),with two gold contacts which will be the source and drain. The device also has a back gate and a top gate which is on dielectric layer (SiO$_2$, PMMA resist or RX resist, for example).  
The back gate controls the charge carrier density of the sample and the top gate modulates the current which flows through the device from source to drain passing through a square potential barrier. We will suppose that the sample is big enough not to be considered as a strip, therefore there will not be edge effects. Furthermore, at temperatures close to 0~K, the charge carrier density $n$ is proportional to Fermi energy squared\cite{Wong} ($n\propto E_F^2$) and will show a linear dependence with the back gate voltage ($n\propto V_{BG}$) \cite{Huard}.

We apply the expressions obtained earlier (Eq. \ref{wfunction})  in order to calculate the transmission coefficient through a square barrier potential $V_0$ and width $D$\cite{Cervero}. Each charge carrier will have an energy $E$ and the angle of incidence $\phi$ will be $-\pi/2<\phi<\pi/2$.\\
The wavefunction will be a linear combination of wavefunctions in regions I and II and III.

\begin{figure}[h]
\centerline{\includegraphics[width = 8.5cm,clip=]{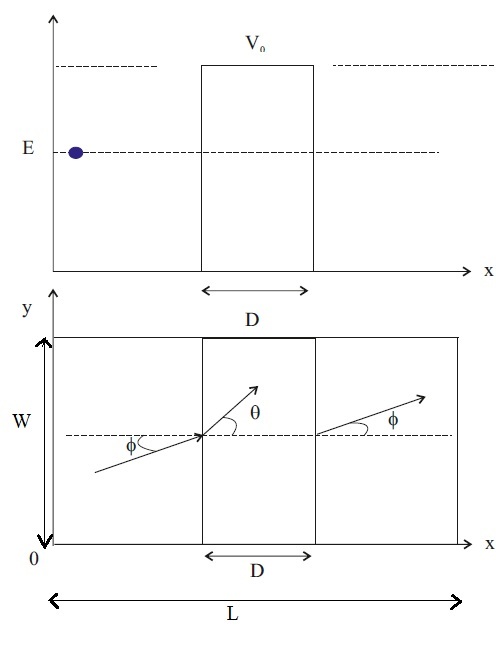}}
\caption{Top panel: Carrier energy and barrier potential dependence with the position in the sample. Bottom panel: scheme of the graphene layer and angles of incidence at the barrier potential}
\label{gauss4}
\end{figure}

\begin{eqnarray}
\Psi_I (x,y)=\frac{1}{\sqrt{2LW}}\left( \begin{array}{clcr}
1\\
s e^{i \phi}\end{array} \right)e^{i(k_x x+k_y y)}+ \nonumber\\
+ 
\frac{r}{\sqrt{2LW}}\left( \begin{array}{clcr}
1\\
s e^{i(\pi-\phi)}\end{array} \right)e^{i(-k_x x+k_y y)}\nonumber\\  
\Psi_{II} (x,y)=\frac{a}{\sqrt{2LW}}\left( \begin{array}{clcr}
1\\
s^{\prime} e^{i \theta}\end{array} \right)e^{i(q_x x+k_y y)} +\nonumber\\
+
\frac{b}{\sqrt{2LW}}\left( \begin{array}{clcr}
1\\
s^{\prime} e^{i(\pi-\theta)}\end{array} \right)e^{i(-q_x x+k_y y)}\nonumber\\  
\Psi_{III} (x,y)=\frac{t}{\sqrt{2LW}}\left( \begin{array}{clcr}
1\\
s e^{i \phi}\end{array} \right)e^{i(k_x x+k_y y)}  
\end{eqnarray}  

For further discussion, we consider that in zone I the amplitude of the incident wavefunction is 1, whereas the amplitude of the reflected wavefunction is $r$. In zone II the amplitude of the incident wavefunction is $a$ and the amplitude of the reflected wavefunction is $b$. Finally, in zone III only transmitted wavefunction with and amplitude of value $t$ is to be observed.
We define $s=sgn(E)$ and $s^\prime=sgn(E-V_0)$ which are related to the nature of the charge carriers (hole or electron regime).

For convenience, we define the following quantities:
$ \phi=\arctan\left(\frac{k_y}{k_x}\right)$,  $k_x = k_F \cos \phi$, $k_y = k_F \sin \phi$,\newline
$k_F=\frac{E}{\hbar v_F}$, $\theta=\arctan\left(\frac{k_y}{q_x}\right)$, $q_x=\sqrt{\left(\frac{V_0 -E}{\hbar v_F}\right)^2 - k_y^2}$.\\

In order to obtain the parameters of the wavefunctions $r$, $a$, $b$, $t$ we will impose boundary conditions such as the wavefunction must be continuous at the borders of the barrier potential. Straightforward algebra, whose details are given in reference \cite{leja}, yields the expression of the transmission coefficient
\begin{widetext}
\begin{equation}
T(E,V_{0},D,\phi)=\left(1+V_{0}^2 tan^2 \phi\frac{ \sin^2 \left(\frac{D}{\hbar v_F} \sqrt{(E-V_{0})^2-E^2 \sin^2 \phi}\right)}{ (E-V_{0})^2-E^2 \sin^2 \phi}\right)^{-1}
\label{ctrans}
\end{equation}
\end{widetext}
As we can see, the transmission coefficient depends upon the energy of the charge carrier $E$, the angle of incidence $\phi$ and the parameters of the barrier potential $V_0$ and $D$ whereas it does not depend upon the dimensions of the graphene sheet $W$ and $L$. Also, the transmission coefficient is symmetric with respect to the angle of incidence $\phi$.
According to this expression, in the case of normal incidence, the transmission coefficient is the unity with independance upon any other parameters. That is the case of the Klein paradox for Dirac particles \cite{Calogeracos},\cite{Katsnelson},\cite{Beenakker}.

The analytical expression obtained for the transmission coefficient  \cite{leja} can be used in order to calculate the conductance across a potential barrier.
The momentum of the charge carriers along the graphene is ${\bf k}=k_x\hat{i}+k_y\hat{j}$. The contribution of each carrier to current density along the x direction is given by:
\begin{eqnarray}\label{Eq2}
j^o_x=- s{e\over LW} &T&(E,V_0,D,\phi) v_F{k_x\over\mid{\bf k}\mid} =\nonumber\\
&=&- s{e\over LW}T(E,V_0,D,\phi)v_F\cos\phi
\end{eqnarray}

The total current is caused by the charges with all possible values of $k_x$ and $k_y$. The number of carriers with momentum between  $( k_x,k_y)$ and $( k_x+dk_x, k_y+dk_y)$ depends on the number of the available states, and its occupation degree, given the Fermi-Dirac distribution  $f_0 (E,\mu) = \displaystyle(1+\exp^ {(E-\mu)/k_BT})^{-1}$.
%for an electron collectivity in thermodynamical equilibrium to a electrochemical potential $\mu$ to a temperature $T$. 

The area element in the momentum space is $\frac{LWdk_xdk_y}{4\pi^2}$ and the number of electrons in this area is (2 due to the spin and 2 due to the duplicity of the valleys of Dirac)

\begin{equation}
dn_{elec}= 2\times 2\times   L W {dk_x dk_y\over 4\pi^2} f_o(E,\mu)=  { L W\over\pi^2} f_o(E,\mu)dk_x dk_y
\end{equation}
Thus, the current density can be calculated by means of
\begin{equation}
dj_x=j^o_x dn_{elec}= - s {e  v_F \over\pi^2}T(E,V_0,D,\phi) f_o(E,\mu)\cos\phi dk_x dk_y
\end{equation} 
We proceed now to use polar coordinates, and use the relation $E=s\hbar v_Fk$
\begin{eqnarray}
dj_x=  -{  4 e\over h^2 v_F }T(E,V_0,D,\phi ) f_o(E,\mu )\cos\phi  E dE d\phi
\end{eqnarray} 
Thus, the total current density is 
\begin{eqnarray}
j_x&=& -{4 e\over h^2 v_F }\int_{-{\pi\over 2}}^{\pi\over 2}\int_{E_1}^{E_2}T(E,V_0,D,\phi )\cdot\nonumber\\
&\cdot& f_o(E,\mu ) E \cos\phi\, dE\, d\phi 
\end{eqnarray} 

If we consider that the current is by both, electrons and holes, $E_1 = - \infty$ and $E_2 = \infty$
\begin{eqnarray}
j_x&=&{-4 e\over h^2 v_F }\int_{-{\pi\over 2}}^{\pi\over 2}\cos\phi\,d\phi\cdot\nonumber\\ 
&\cdot&\int_{-\infty}^{\infty}T(E,V_0,D,\phi ) f_o(E,\mu )E\,dE
\end{eqnarray}

We consider now that the contacts are actually nothing but ideal reservoirs which stablish equilibrium of the distribution of electrons to a certain chemical potential $\mu_L$ (source) and $\mu_R$ (drain) yielding two electron fluxes. When these two currents are considered, being symmetric the transmission coefficient as it is, we obtain:

\begin{eqnarray}
j_x&=& - { 4 e\over h^2 v_F }\int_{-{\pi\over 2}}^{\pi\over 2}\cos\phi\,d\phi\int_{-\infty}^{\infty}T(E,V_0,D,\phi ) \Big(f_o(E,\mu_{L} )-\nonumber\\
&-&f_o(E,\mu_{R} )\Big)E\, dE
\end{eqnarray} 

The later expression can be reduced in some cases :
\begin{itemize}
\item If the bias that exists is large at low temperature, the contribution due to the reservoir with lesser chemical potential (right) is negligible. It is possible to obtain the total density of current if we consider only the current that provide the reservoir on the left.

\item At low temperature ($k_B T<E_F-\mu$), when the electrons are hightly degenerated, the Fermi function can be approximated by the step function, and therefore
\begin{equation}
j_x={- 4 e\over h^2 v_F }\int_{-{\pi\over 2}}^{\pi\over 2}\cos\phi d\phi \int_{\mu_R}^{\mu_L}T(E,V_0,D,\phi ) E dE
\end{equation}

\item  If the bias is very small, we can approximate the Fermi distribution by its Taylor approximation around the mean value of the chemical potentials $\mu = {\mu_L+\mu_R\over 2}$  and considering that the difference between the chemical potentials is established in fact by means of the bias $ \mu_L -\mu_R = e\,(V_R -V_L)$:
\begin{eqnarray}
f_o(E,\mu_L )-f_o(E,\mu_R )&\approx& {\partial f_o\over\partial\mu} (\mu_L-\mu_R)=\nonumber\\
&=&- e {\partial f_o(E,\mu )\over\partial E} (V_R-V_L)
\end{eqnarray}
and therefore

\begin{eqnarray}
j_x&=& - { 4 e^2\over h^2 v_F } (V_R-V_L)\int_{-{\pi\over 2}}^{\pi\over 2}\cos\phi\,d\phi\cdot\nonumber\\
&\cdot&\int_{-\infty}^{\infty}T(E,V_0,D,\phi ) \Big (- {\partial f_o(E,\mu )\over\partial E} \Big ) E\,dE   \nonumber
\end{eqnarray}

The minus sign describes the correct direction of the electrical current carried for both electrons and holes. Therefore, since the conductance is given by $G=\frac{I}{V}$ and the intensity is $I=\int_{0}^{W}j_xdy$  the conductance is given by $G={j_x\over \Delta V}W$. We define the effective conductance as $G_{eff}=G/W$, and
\begin{eqnarray}
 G_{eff}&\approx &{ 4 e^2\over h^2 v_F }\int_{-{\pi\over 2}}^{\pi\over 2}\cos\phi\,d\phi\times\nonumber\\
&\times&\int_{-\infty}^{\infty}T(E,V_0,D,\phi ) 
\Big (- {\partial f_o(E,\mu )\over\partial E} \Big ) E\,dE
\end{eqnarray} 
\item At ultra low temperatures (at zero temperature $\mu=E_{F}$) we can approximate 
\begin{equation}
- {\partial f_o(E,\mu )\over\partial E}\approx{\delta ({E-\mu})}={\delta ({E-E_{F}})}
\end{equation} 
so that we obtain
\begin{equation}
 G_{eff}\approx{ 4 e^2\over h^2 v_F } |E_{F}|\int_{-{\pi\over 2}}^{\pi\over 2} T(E_{F},V_0,D,\phi ) \cos\phi\,d\phi
\end{equation}
 In units of $e^2/h$:
\begin{equation}\label{Geff}
 G_{eff}=\frac{2}{\hbar\,v_{F}\,\pi}|E_{F}|\int_{ -\pi/2}^{ +\pi/2} T(E_{F},V_0,D,\phi)\cos\phi\,d\phi             
\end{equation}
\end{itemize} 
We point out that if we express the lengths in {\it nm} and the energies in  {\it meV}, it results $\hbar v_F=658.2$ $meV\,nm$  and  the effective conductance is given in units of $e^2/h$.

The conductance will depend on the potential applied by the top gate (through the transmission coefficient), the potential applied by the back gate (through $E_F$) and the width of the graphene sheet, but not upon its length.

Furthermore, we observe in (Eq.\ref{Geff}) that the maximun value of the effective conductance is at $T(E_{F},V_0,D,\phi)=1$, and therefore
\begin{equation}\label{Gmax}
 G_{eff,max}=\frac{4}{\hbar\,v_{F}\, \pi}|E_{F}|  
\end{equation}

By introducing the expresion in (Eq.\ref{ctrans})  into (Eq.\ref{Geff}) we obtain the effective conductance for one square barrier potential in graphene:

\begin{widetext}
 \begin{equation}\label{Gb}
G_{eff}(E_{F},V_{0},D)=\frac{2\,|E_{F}|}{\hbar\,v_{F}\, \pi}\int_{ -\pi/2}^{ +\pi/2}\left(1+V_{0}^2 tan^2 \phi\frac{\sin^2 \left(\displaystyle\frac{D}{\hbar\, v_F} \sqrt{(E_{F}-V_{0})^2-E_{F}^2 \sin^2 \phi}\right)}{(E_{F}-V_{0})^2-E_{F}^2 \sin^2 \phi}\right)^{-1}\, \cos\phi\,d\phi        
\end{equation}
\end{widetext}

As it is seen, the effective conductance of a graphene-based FET depends upon the Fermi energy and the properties of the potential barrier created by the top gate  (height $V_0$ and width $D$) .
\begin{figure}
\centering
\includegraphics[width = 8.5cm,clip=]{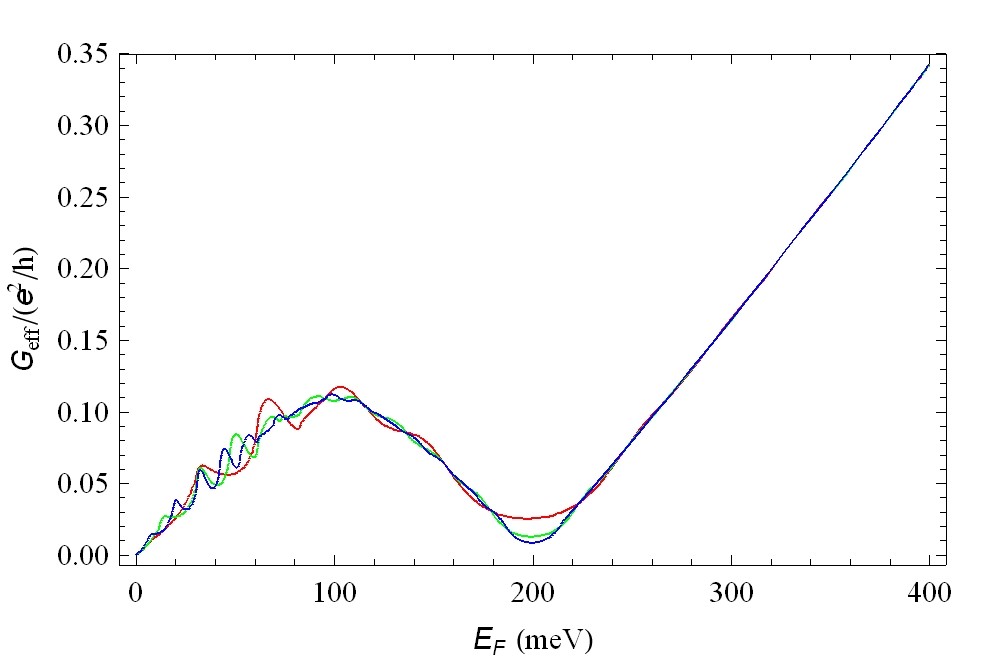}
\caption{Effective conductance versus Fermi energy at height of the barrier $V_{0}$ = 200 meV and width of the barrier $D$ 50(red), 100(green) and 150(blue) nm. Conductance has a local maximum at $E_F=0.5V_0$, a local minimum at $E_F=V_0$ and increases linearly with $E_F\geq V_0$.}
\label{f1}
\end{figure}

Figure \ref{f1} shows conductance versus Fermi energy. It is shown that the curve of the effective conductance has a local maximum when $E_{F} = 0.5V_{0}$.  At this point, the sign of the quantity $ (E_{F}-V_{0})^2-E_{F}^2 \sin^2 \phi$ is changed from positive to negative. Therefore,  in (Eq.\ref{Gb}) the sine of this quantity will become a hyperbolic sine, and the conductance will become smaller. Thus, the resistance will increase. This situation will persist until a local minimum is reached, at  $E_{F} \simeq V_{0}$. When $E_{F}$ is greater than $V_{0}$, since the transmission coefficient tends to unity, the effective conductance grows proportionally to the Fermi energy, similarly to (Eq.\ref{Gmax}). We observe oscillations of the conductance at Fermi energies which verify $E_{F}<V_{0}$. For  $E_{F}<0.5V_{0}$ there are several oscillations, whereas  for values of the Fermi energy such that $0.5V_{0}<E_{F}<V_{0}$ there is only one oscillation. If $E_{F}>>V_{0}$ the conductance does not depend on the value of the barrier width and varies linearly with the Fermi energy.

\begin{figure}
\centering
\includegraphics[width = 8.5cm,clip=]{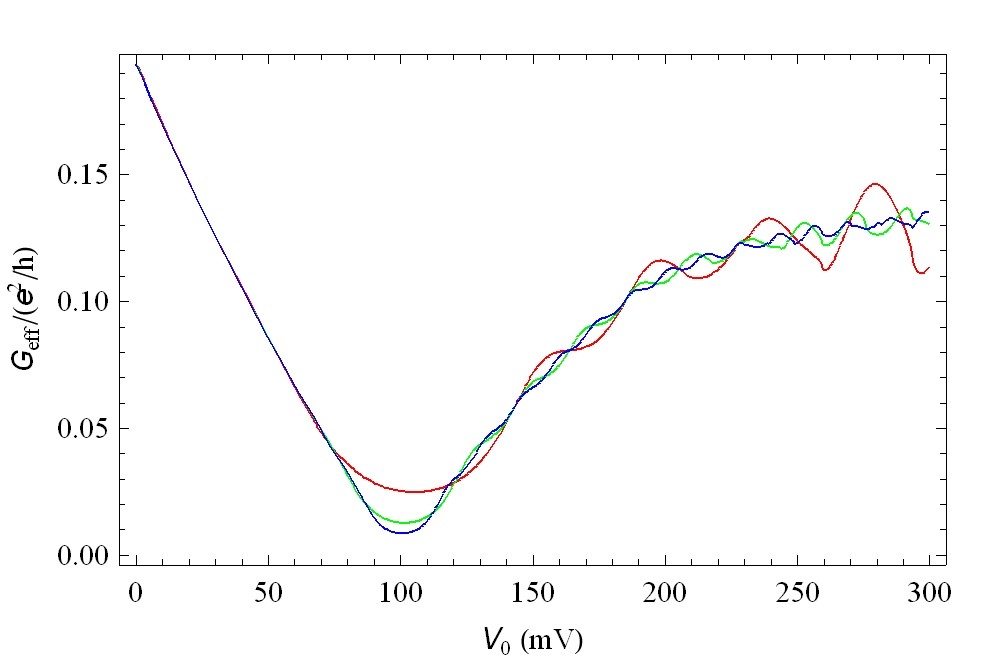}
\caption{Conductance versus height of the barrier at Fermi energies $E_{F}$ = 100 meV and $D$ 50(red), 100(green) and 150(blue) nm. The three curves take the same values for $E_F>V_0$, and present a minimum for $E_F=V_0$. The amplitude of the oscillations depends on the value of the width of the barrier $D$.}
\label{f1a}
\end{figure}

As we can see in figure \ref{f1a}; when $V_{0}$ increases, the effective conductance decreases linearly and takes the same value indepently of the value of the width, for $V_{0}$ verifying $V_{0}< E_{F}$. When $V_0$ takes the value of the Fermi energy, there is a local minimum, whose value depends on the width of the barrier $D$. For $V_0>E_F$, the conductance increases and exhibits oscillations which are bigger as the width of the barrier decreases. Furthermore, it is appreciated that, as the value of $D$  decreases the amplitude of the oscillations increases. These oscillations appear only for values of $V_0$ which verify $V_0>E_{F}$.

\begin{figure}[hbtp]
\centering
\includegraphics[width = 8.5cm,clip=]{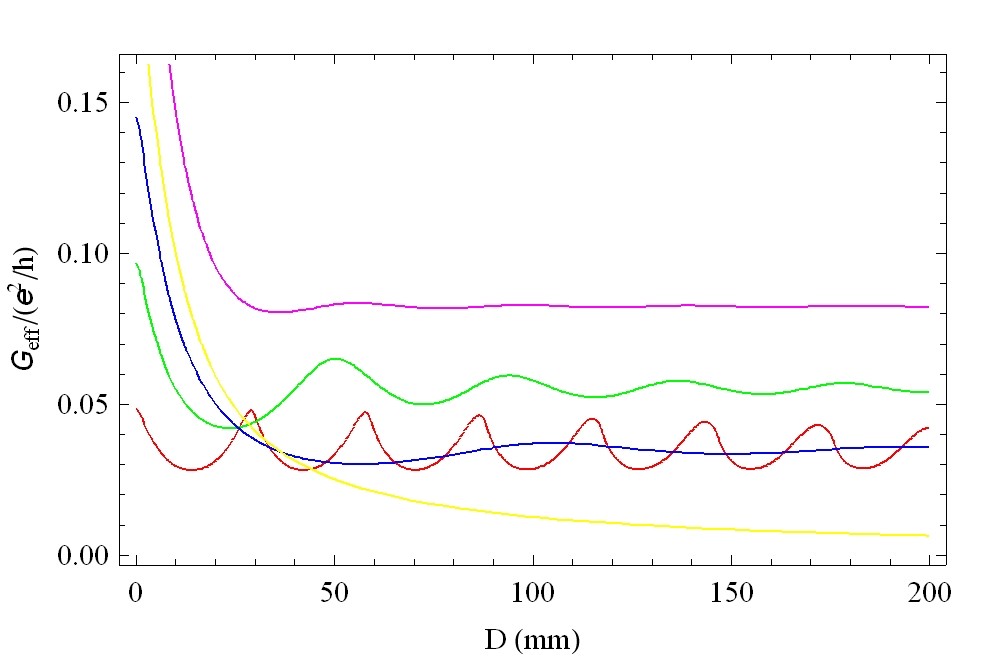}
\caption{Conductance versus width of the barrier when assuming a height of the barrier $V_0$ = 100 meV and Fermi energies $E_{F} = $ 25(red), 50(green), 75(blue), 100(yellow) and 150(purple) meV in order to perform the calculations.}
\label{fd}
\end{figure}

The dependance of the effective conductance upon the width $D$ of the barrier is shown in figure \ref{fd}: the conductance decreases until it tends to be stabilised in the vicinities of a constant value. It is also observed that the oscillations tend to be lesser as the Fermi energy grows.

\section{Weighted effective conductance}

Up to now we have considered that the carriers scatter through the barrier with the same probability for each angle of incidence. It should be considered another scenarios in which, due to the inhomogeneities of the material or the design of the device, the carriers flow around certain direction. Thus, the angle of incidence will be different from 0$^\circ$ and will obey certain probability distribution function $P(\phi)$ centered around an angle $\phi_0$.
This probability distribution function must verify the condition

\begin{equation}\label{norm}
\int_{ -\pi/2}^{ +\pi/2} P(\phi)d\phi=\pi
\end{equation}

Next, we will consider different probabilities of distribution and we will study its effect on the effective conductance.

\subsection{Gaussian distribution}

We consider now the case of a Gaussian function for the probability distribution of the angle of incidence. This assumption represents that transport is of ballistic nature and that there are a few inhomogeneities which cause scattering processes within the sample and deviates carriers from their ballistic trajectory; making them to impact the barrier potential with a non zero angle of incidence. Since not all the carriers will suffer the same scattering process, we can assume that the majority of carriers will impact the barrier potential with a similar angle $\phi_0$ and that there will be more carriers impacting the barrier potential with an angle of incidence $\phi_0$ than carriers impacting the barrier with an angle of incidence that differs substantially from $\phi_0$. 
Thus, we assume that the angles of incidence of the carriers will obey  a probability distribution function that has the form $P(\phi)=c\exp\left(-\frac{\left(\phi-\phi_0\right)^2}{2a^2}\right)$.\\

The parameter $a$ of the Gaussian functions makes the distinction between an homogeneus sample (a narrow Gaussian bell) or a sample with plenty of inhomogeneities (a wide distribution).
A narrow Gaussian function represents the case in which almost all the carriers scatter with the barrier at the same angle of incidence. This will happen when the sample is homogeneus and the carriers do not scatter with the sample inhomogeneities, and therefore the transport will be ballistic and almost all the carriers will move through the sample with a similar direction.
Obviously, a wide Gaussian function represents the opposite case.\\

We start the analysis by considering the scenario in which the transmission coefficient equals the unity: $\phi_0$ =0$^\circ$.\\

Obviously, the parameters $a$ and $c$ are not independent, since they must hold the normalization condition $ac\sqrt{2\pi}Erf\left(\frac{\pi}{2\sqrt{2}a}\right)=\pi$. As we can see, as $c$ becomes smaller $a$ becomes greater.

The conductance is given by:

\begin{widetext}
\begin{equation}
 G_{g}(E_{F},V_{0},D)=\frac{2\,c}{\hbar\,v_{F}\, \pi}|E_{F}|\int_{ -\pi/2}^{ +\pi/2} T(E_{F},V_{0},D,\phi)\cos\phi \, \exp\left(-\frac{\phi^2}{2a^2}\right) d\phi             
\end{equation}
\end{widetext}

\begin{figure}
\centering
\includegraphics[width = 8.5cm,clip=]{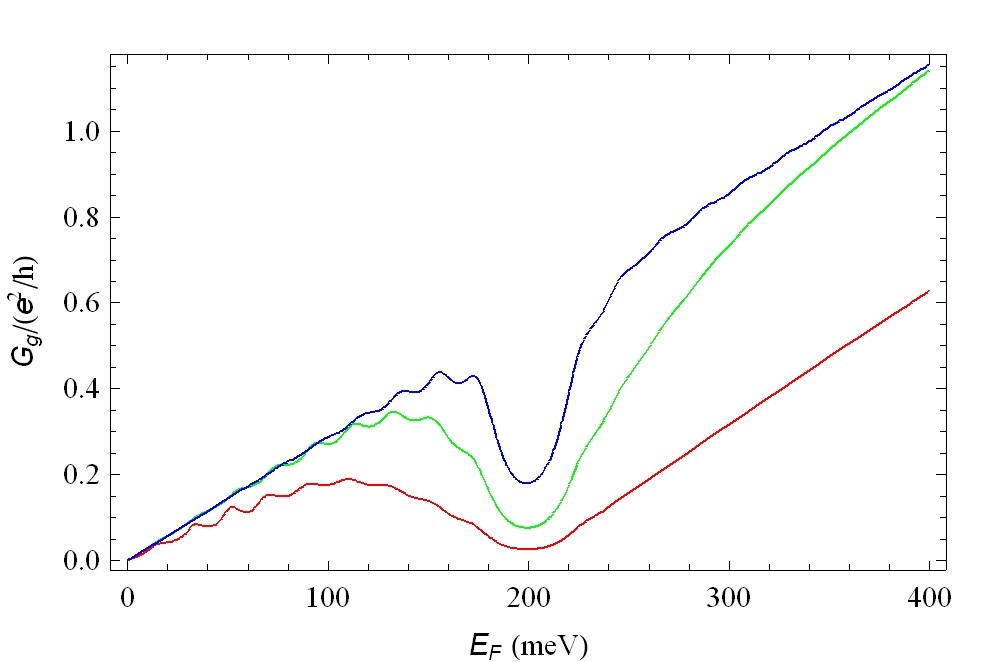}
\caption{Effective conductance versus the Fermi energy when the distribution function of the angle of incidence is a Gaussian function centered around $\phi_0$ = 0$^\circ$ assuming different widths of the Gaussian: 0.64 (red), 0.21 (green) and 0.08 (blue). The height of the barrier is $V_{0} = 200 $~meV and its width is $D = 100$~nm.}
\label{gauss3}
\end{figure}

\begin{figure}
\centering
\includegraphics[width = 8.5cm,clip=]{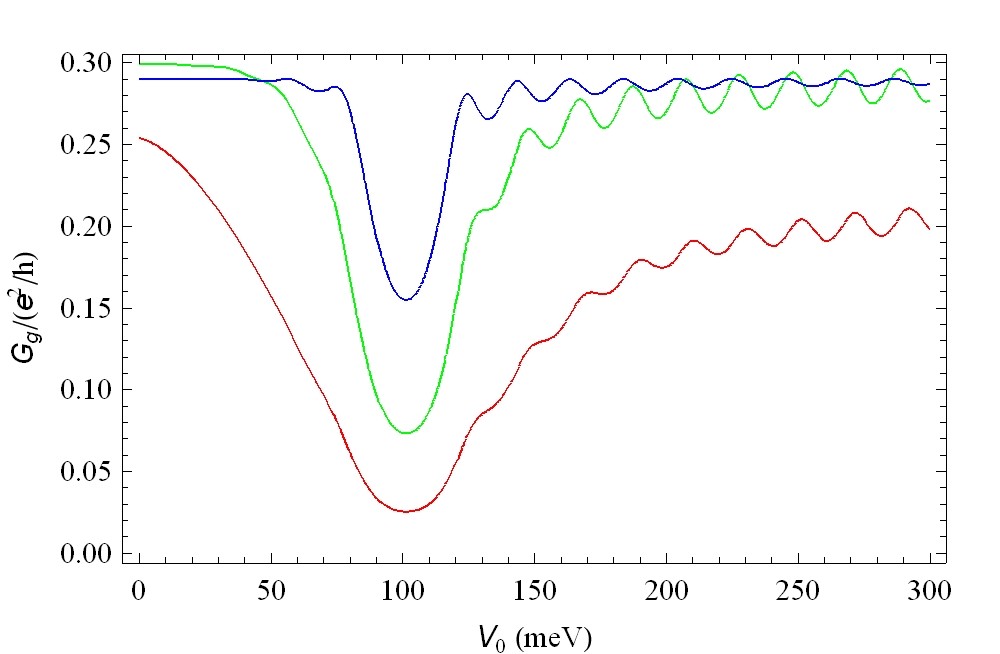}
\caption{Conductance versus the height of the barrier $V_0$ asuming a Gaussian probability distribution function centered around $\phi_0$ = 0$^\circ$. The width of the Gaussian distribution is $a=0.64$ (red), $a= 0.21$ (green) and $a=0.08$ (blue). The Fermi energy of the system is assumed to be for $ E_F= 100 $~meV and the width of the barrier is $D = 100$~nm.}
\label{gauss2}
\end{figure}

As in the previous case, the effective conductance depends on $E_{F}$, $ V_{0}$ and  $D$ but it shows a different behaviour, as shown in figures \ref{gauss3} and \ref{gauss2}.\\

We study the behaviour of the conductance when the Fermi energy is changed while the other variables remain at fixed values (see figure~\ref{gauss3}).
The conductance oscillates whilst it increases until a local maximum is reached. The value and position of this maximum depends on the parameters of the Gaussian function used. Once the maximum is reached, the conductance decreases until it reaches a local minimum when $E_F\sim V_0$. Once the minimum is surpassed, the conductance increases monotonically.\\

At it is shown in figure \ref{gauss2}, when using $V_0$ as the driving parameter and the other parameters remain fixed at constant values, we can see that the effective conductance decreases until it reaches a minimum around $E_F\sim V_0$. It is also observed that the narrower Gaussian function (which is equivalent to a less disordered distribution) the variation of the conductance is more pronounced. Once $E_F$ is greater than $V_0$ we can observe oscillations of the conductance whose frequency tends to the value $\omega=2D/\hbar v_F$. \\

\begin{figure}[hbtp]
\centering
\includegraphics[width = 8.5cm,clip=]{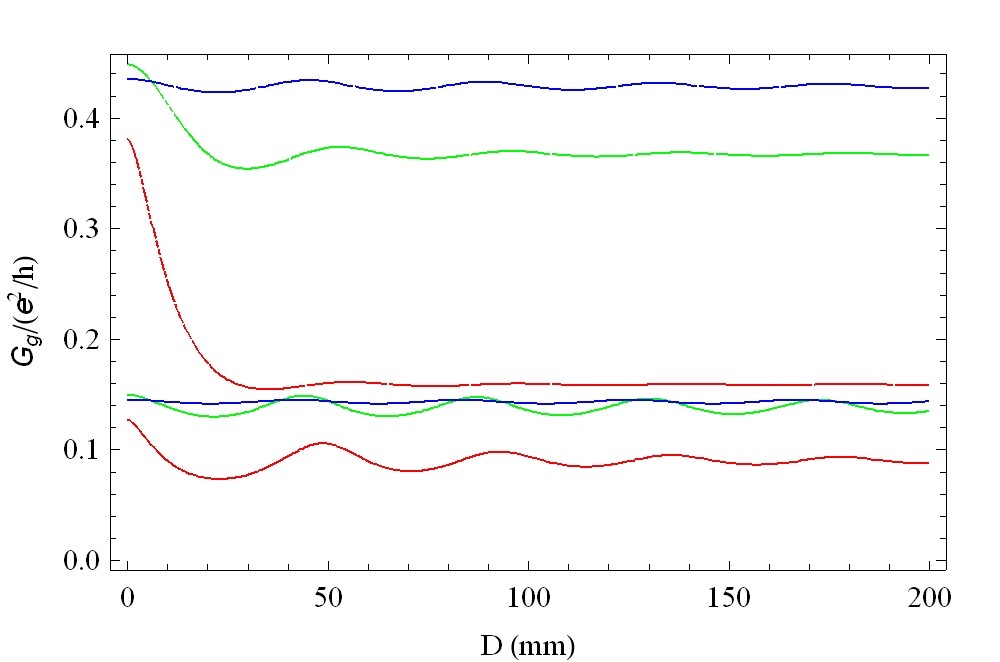}
\caption{Conductance versus the width of the barrier using a Gaussian distribution of probability centered around $\phi_0$ = 0$^\circ$. The width of the Gaussian is assumed to be $a=0.64$ (red),$a= 0.21$ (green) and $a=0.08$ (blue). The Fermi energy is $ E_F = 50$~meV (line) and  $E_F=150$~meV(dashed line). Also, the height of the barrier is $ V_{0} = 100$~ meV.}
\label{gaussd}
\end{figure}

We consider now the case in which the Gaussian distribution is not centered around $\phi_0=0^\circ$. This might be done by tilting the orientation of the top gate in a graphene based device as, for example, shown in \cite{Low}.
In this case, the normalization condition is given by the expression $ac\sqrt{\frac{\pi}{2}}\left(Erf\left(\frac{-2 \phi_0 + \pi}{2\sqrt{2}\cdot a}\right) +Erf\left(\frac{2 \phi_0 + \pi}{2\sqrt{2}\cdot a}\right)\right) =\pi$.

\begin{figure}
\centering
\includegraphics[width = 8.5cm,clip=]{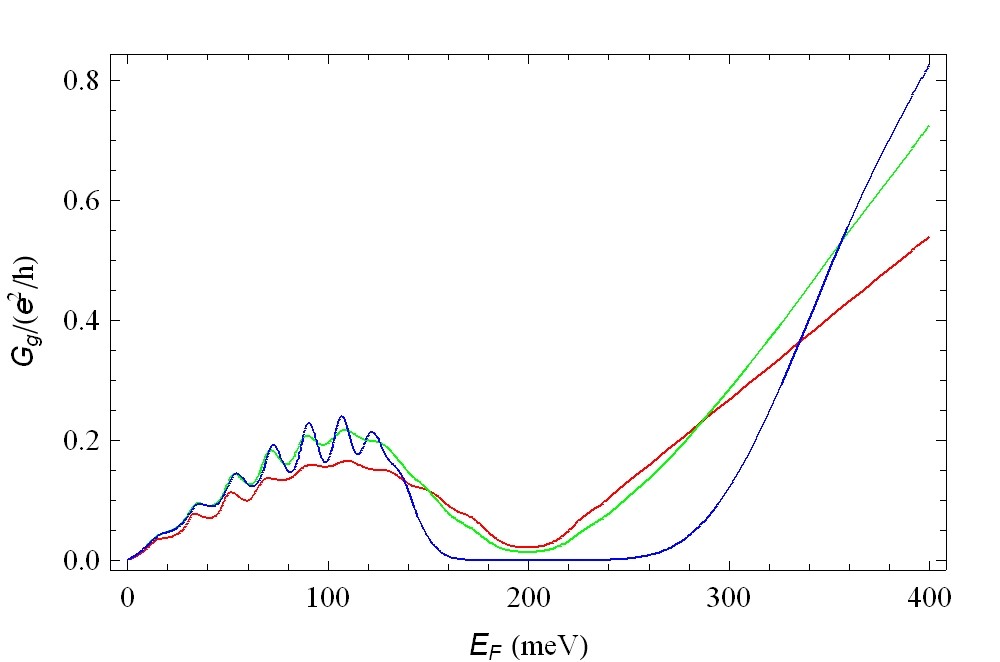}
\caption{Conductance versus the Fermi energy considering that the distribution is a Gaussian function centered around $\phi_0$ = 22.5$^\circ$ using different widths for the Gaussian function: $a=0.65$ (red), $a=0.21$ (green), $a=0.08$ (blue). The height of the barrier is $V_{0} = 200$~ meV and its width  $D = 100$~ nm.}
\label{gauss31}
\end{figure}

\begin{figure}[hbtp]
\centering
\includegraphics[width = 8.5cm,clip=]{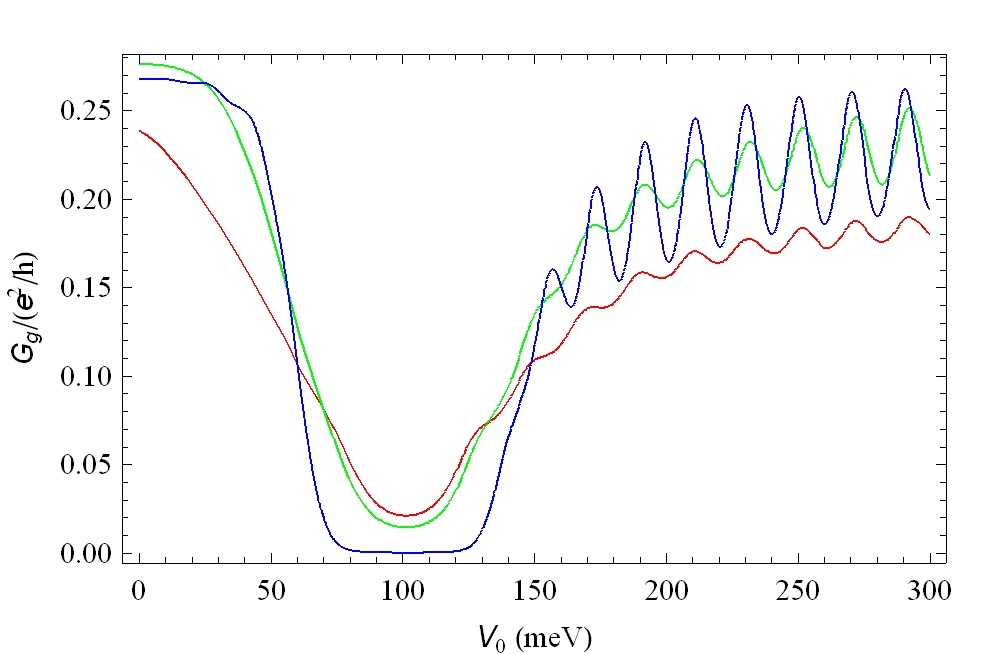}
\caption{Effective conductance versus the height of the barrier $V_0$ considering that the distribution is a Gaussian function centered around $\phi_0$ = 22.5$^\circ$  using different widths for the Gaussian function: $a=0.65$ (red), $a=0.21$ (green), $a=0.08$ (blue). The Fermi energy is $E_F=100$~meV and its width  $D = 100$~ nm. We note that as the Gaussian becomes narrower, \emph{i.e,} as the number of inhomogeneities is decreased, the minimum of conductance is wider.}
\label{gauss21}
\end{figure}

As we can see in figures \ref{gauss31} and \ref{gauss21} the results are similar than the obtained in the previous case, but the minima observed previously are wider. In fact as $a$ is closer to zero these minima become wider.

\begin{figure}[hbtp]
\centering
\includegraphics[width = 8.5cm,clip=]{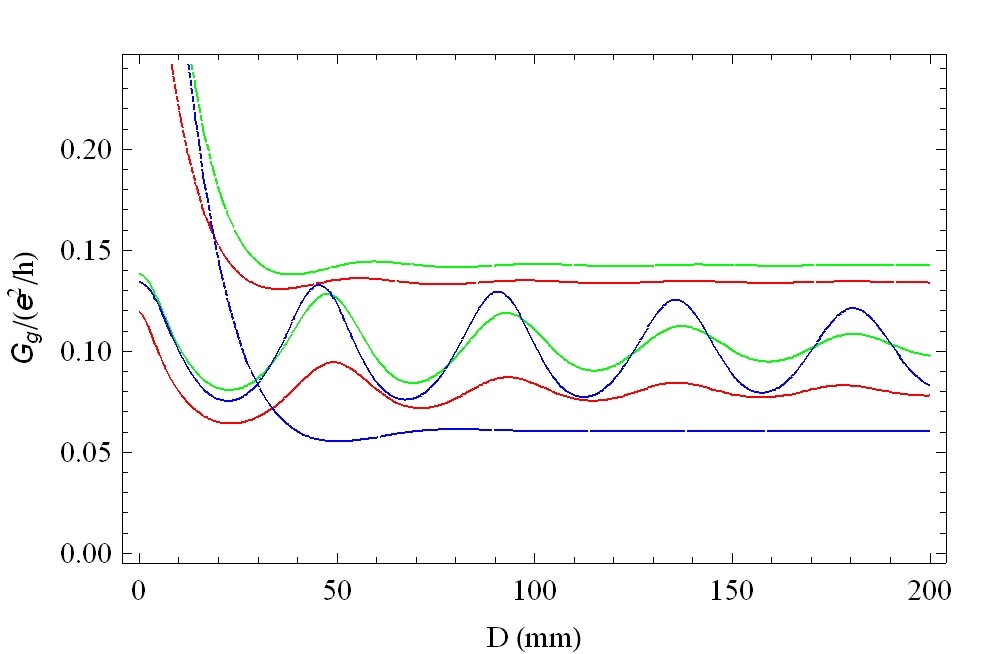}
\caption{Effective conductance versus the width of the barrier $D$ assuming that the distribution is a Gaussian function centered around $\phi_0$ = 22.5$^\circ$ for $a=0.65$ (red), $a=0.21$ (green), $a=0.08$ (blue) for $ E_F$ = 50(line), 150 meV(dashed line) and $ V_{0}$ = 100 meV.}
\label{gaussd1}
\end{figure}

\subsection{Delta distribution}

Next, we consider the extreme scenario in which the Gaussian function is so narrow that it can be represented by a Delta function $\delta(\phi-\phi_0)$. This is the case in which the material is ideal and all the carriers scatter through the barrier with the same angle of incidence $\phi_0$.  

In this case, the conductance is given by:

\begin{equation}
G_{d,\phi_{0}}(E_{F},V_{0},D)=\frac{2\,|E_F|}{\hbar\,v_{F}}  T(E_{F},V_{0},D,\phi_{0})\cos\phi_{0} 
\end{equation}

It is observed that the effective conductance depends now on the angle of incidence, along with the rest of the parameters previously considered (Fermi energy, $E_F$, height of the barrier, $V_0$, and width of the barrier, $D$).\\

\begin{figure}
\centering
\includegraphics[width = 8.5cm,clip=]{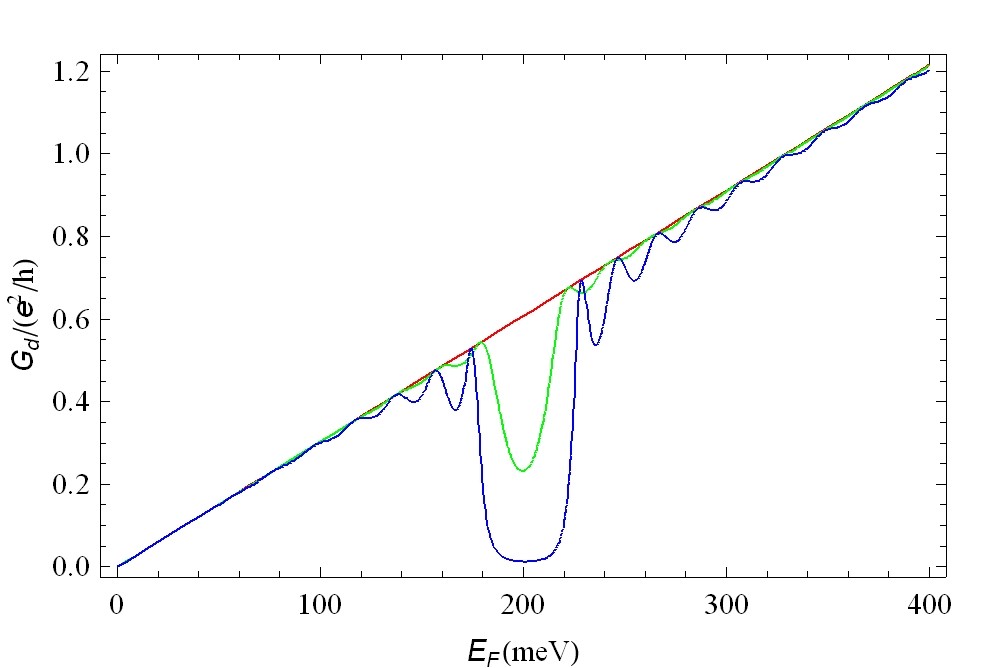}
\caption{Effective conductance versus Fermi energy assuming a delta like distribution when the barrier is assumed to be such that its height is $V_{0} = 200$~ meV and its width $D = 100$~ nm. We assume different angles of incidence $\phi_0 =$  0$^\circ$(red), 2$^\circ$(green) and 5$^\circ$(blue). These curves are the limit of a narrow gaussian.}
\label{gdelta2}
\end{figure}

In particular, if the current flows perpendicularly to the barrier, the transmission coefficient equals 1 with no dependance on the Fermi energy and the parameters of the barrier, as a consequence of the Klein paradox\cite{Calogeracos},\cite{Katsnelson},\cite{Beenakker}. Therefore, the conductance is proportional to the Fermi energy: $$ G_{d,0}=\frac{2}{\hbar\,v_{F}}|E_{F}|$$
For non normal incidence, the conductance presents an effective gap between the values ${V_0\over 1\pm\sin\phi_0}$ ​​of the  Fermi energy,  developing  an effective minimum at $E_F = V_0/\cos^2 \phi_0$ (as shown in figure \ref{gdelta2}), whose value is
$$G_{min} = {2\,V_0\over\hbar\,v_F}\cosh^{-2}\left( { DV_0\tan\phi_0\over\hbar\,v_F}\right) \cos^{-1}\phi_0$$

\begin{figure}
\centering
\includegraphics[width = 8.5cm,clip=]{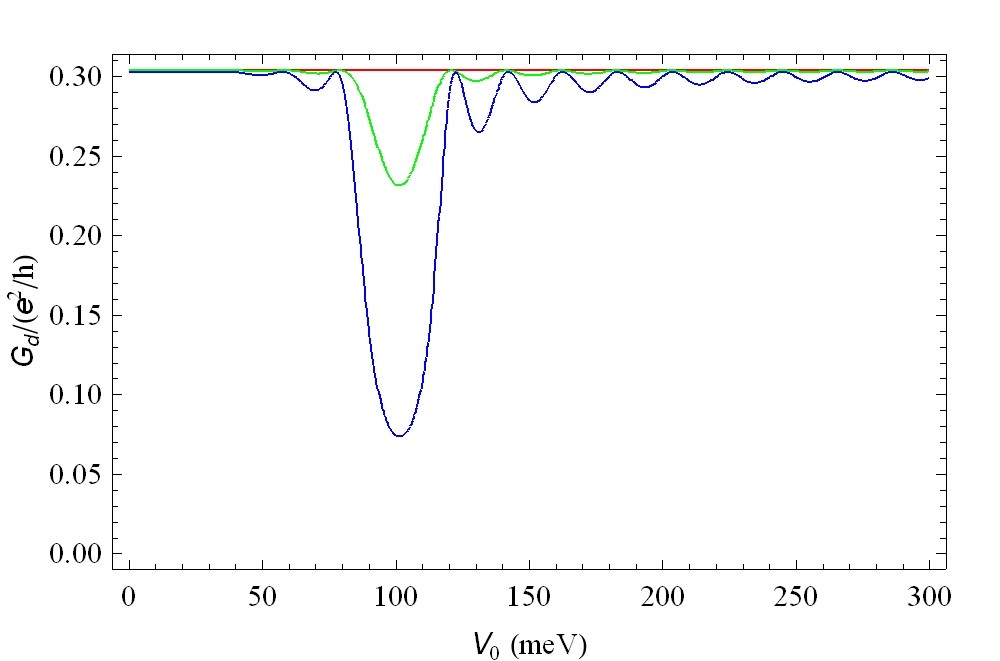}
\caption{Conductance versus the height of the barrier assuming a delta like distribution supposing that the Fermi energy is $E_F$ = 100 meV and that the width of the barrier is $D$ = 100~nm. This calculation has been performed assuming different angles of incidence $\phi_0 =$  0$^\circ$(red), 2$^\circ$(green) and 5$^\circ$(blue).}
\label{gdelta1}
\end{figure}

\begin{figure}[hbtp]
\centering
\includegraphics[width = 8.5cm,clip=]{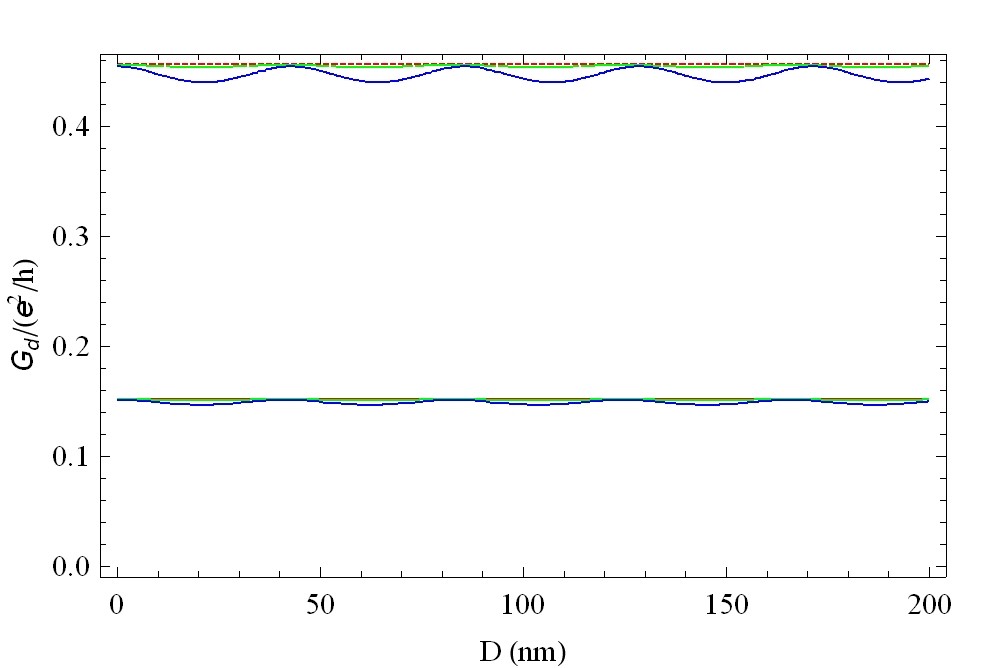}
\caption{Effective conductance versus the width of the barrier if a delta like distribution probability is assumed at Fermi energies $E_{F}$ = 50(line), 150 meV(dashed line), and the height of the barrier potential is $V_0$ = 100 meV. We have assumed different angles of incidence: $\phi_0 =$  0$^\circ$(red), 2$^\circ$(green) and 5$^\circ$(blue).}
\label{gdeltad}
\end{figure}

\begin{figure}[hbtp]
\centering
\includegraphics[width = 8.5cm,clip=]{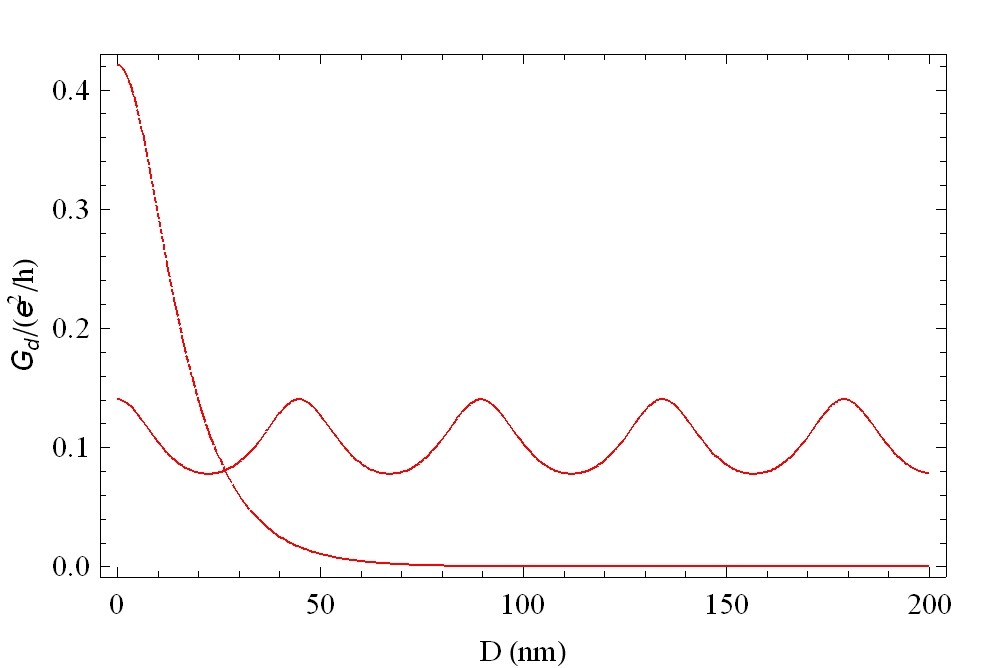}
\caption{Conductance versus the width of the barrier assuming a delta like function centered around  $\phi_0$ = 22.5$^\circ$ for $E_{F}$ = 50(line), 150 meV(dashed line) and $V_o$ = 100 meV.}
\label{gdeltad1}
\end{figure}

In this scenario in which we assume a delta-like probability distribution function, we can see that for an angle of incidence of $\phi_0=22.5^\circ$ (or $\pi/8$) a wide minimum of the conductance is developed (figures \ref{gdelta2} and \ref{gdelta1}). This minimum is formed both when using the Fermi energy (which can be associated with the back gate voltage of the transistor shown in  figure \ref{c3}) and the height of the barrier (the top gate voltage in our model schematized) as a driving parameter. In this case, the dependance of the conductance on the width of the barrier shows no remarkable features but a small oscillation whose amplitude depends on the Fermi energy (see figures \ref{gdeltad} and \ref{gdeltad1}).
 
\begin{figure}
\centering
\includegraphics[width = 8.5cm,clip=]{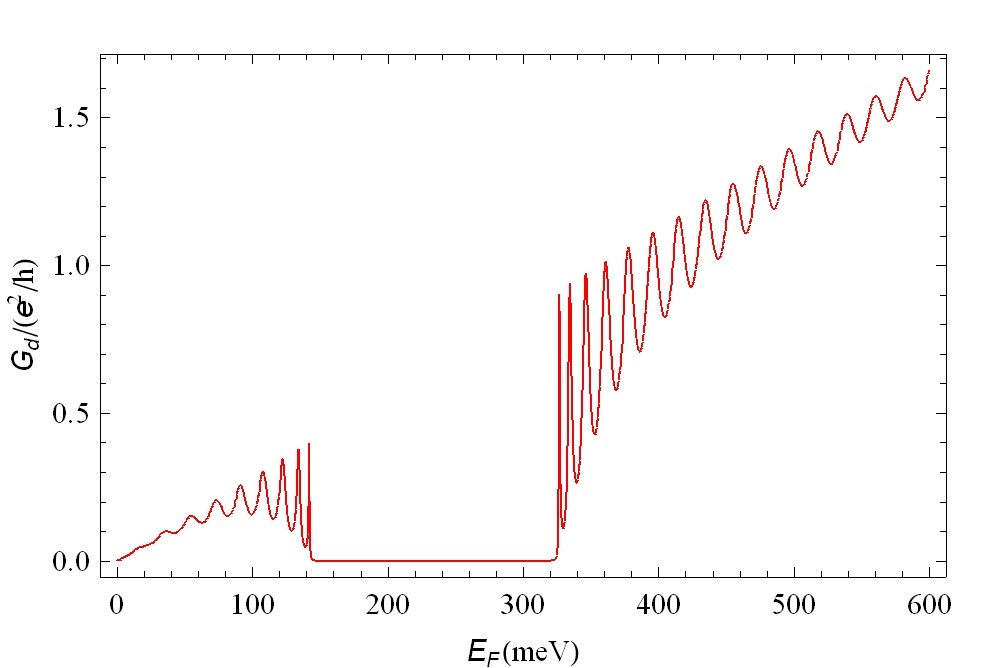}
\caption{Effective conductance versus the Fermi energy when a delta-like probability distribution function is assumed at angle of incidence $\phi_0$ = 22.5$^\circ$ and when the potential barrier is such that $V_{0}$ = 200 meV and $D$ = 100 nm.}
\label{gdelta21}
\end{figure}

\begin{figure}
\centering
\includegraphics[width = 8.5cm,clip=]{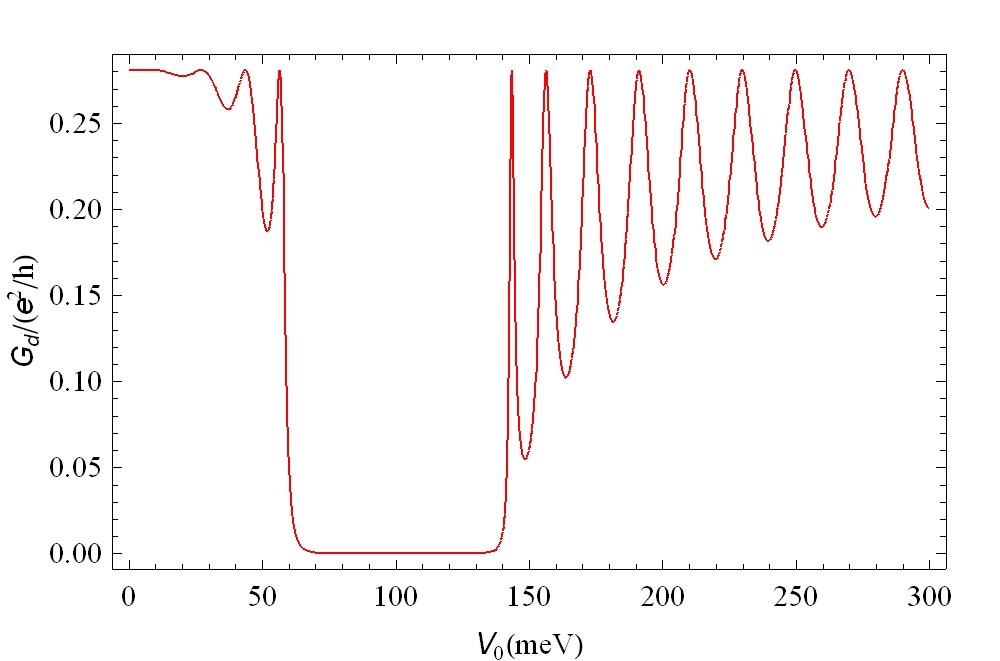}
\caption{Conductance versus the height of the barrier $V_0$ when a delta like probability distribution function is assumed at angle of incidence $\phi_0$ = 22.5$^\circ$ and the other parameters are such that $E_F$ = 100 meV and $D$ = 100 nm.}
\label{gdelta11}
\end{figure}

As the angle of incidence is increased, the features described tend to be greater. In particular, the effective gap which appears is wider and more pronounced, as seen in figures \ref{gdelta21} and \ref{gdelta11} where we have assumed an angle of incidence $\phi_0=22.5^\circ$.
The observed minimum of the conductance shows that in certain conditions the Klein tunneling might be avoided and that the fabrication of an effective FET based on graphene with a noticeable on-off ratio is feasible.

\section{Conclusion}

We have obtained an analytical expression of the transmission coefficient through a square barrier potential based on the previous work done \cite{leja} in which the continuity conditions were applied in order to obtain the Dirac functions of carriers in graphene in the presence of a barrier potential.\\

Within this approach, we have stablished a theoretical model to study transport through a square barrier potential in graphene. Analytical expressions for the transmission coefficient in several different scenarios have been obtained. These different scenarios are a modelization of the different scattering conditions of the sample due to the different degree of inhomogeneity in the sample.

In this model we have also studied a graphene based device and its  effective conductance between its two terminals. In particular, we have found an scenario in which the Klein paradox is neglected and the creation of an effective gap is induced. Therefore we think that it might be suitable to control the transport in a graphene based field effect transistor.\\

We believe that a top gate which is non-perpendicular to the source-drain direction might create a gap in the case in which the graphene is clean and shows almost no inhomogeneities. This phenomenon occurs for any angle of incidence $\phi_0$. The position and width of the energy gap, depend on both the barrier height and $\phi_0$ itself. This feature might be an important step in the development of transistors based on graphene, as the non existance of a gap is one of the main difficulties found to effectively modulate the current via the bias.\\

The authors thank Jos\'{e} Mar\'{\i}a Cerver\'{o} for useful discussion. The authors greatly acknowledge the financial
support of this research from the DGICYT under project FIS2009-07880 and JCYL SA049A10-2 and MEC through grant AP2009-2619

\end{document}